\documentclass[aps, prb, reprint, superscriptaddress]{revtex4-2}

\usepackage[utf8]{inputenc}
\usepackage[T1]{fontenc}
\usepackage{newtxtext}
\usepackage{newtxmath}
\usepackage[scaled=0.92]{helvet}
\usepackage[english]{babel}
\usepackage{microtype}
\usepackage{xspace}
\usepackage{amsmath, amsfonts}
\usepackage{bm}
\renewcommand{\epsilon}{\varepsilon}
\renewcommand{\theta}{\vartheta}
\renewcommand{\rho}{\varrho}
\renewcommand{\phi}{\varphi}
\newcommand*{\abs}[1]{\left\lvert #1 \right\rvert}

\renewcommand{\vec}[1]{\bm{#1}}
\newcommand*{\hamil}{\mathcal{H}}
\DeclareMathOperator\tr{tr}
\usepackage[range-units=single, list-units=single, detect-all]{siunitx}
\DeclareSIUnit\rydberg{Ry}
\usepackage{graphicx}
\usepackage[autostyle]{csquotes}
\usepackage{booktabs}
\usepackage[breaklinks, hidelinks]{hyperref}

\newcommand*{\crystaldir}[1]{$[#1]$}

\providecommand*{\tab}{Tab.\@\xspace}
\providecommand*{\fig}{Fig.\@\xspace}
\providecommand*{\FFig}{Figure\@\xspace} 
\providecommand*{\ref}{Ref.\@\xspace}

\providecommand*{\eq}{Eq.\@\xspace}

\graphicspath{{graphics/}} 
\newcommand*{\acronym}[1]{\textsc{\MakeLowercase{#1}}} 
\newcommand*{\fremdwort}[1]{\emph{#1}} 
\newcommand*{\tensorial}[1]{\mathcal{#1}}

\newcommand*{\affilUKN}{
  Fachbereich Physik, 
  Universität Konstanz, 
  D-78457 Konstanz, 
  Germany} 
\newcommand*{\affilBME}{
  Department of Theoretical Physics, Institute of Physics, 
  Budapest University of Technology and Economics, 
  Műegyetem rkp.\ 3., 
  H-1111 Budapest, 
  Hungary} 
\newcommand*{\affilJGU}{
  Institute of Physics, 
  Johannes Gutenberg University Mainz, 
  Staudingerweg 7, 
  55128 Mainz, 
  Germany} 
\newcommand*{\affilMTA}{
  ELKH-BME Condensed Matter Research Group,  
  Budapest University of Technology and Economics, 
  Műegyetem rkp.\ 3., 
  H-1111 Budapest, 
  Hungary} 
\newcommand*{\affilWRC}{
  Department of Theoretical Solid State Physics,  
  Institute for Solid State Physics and Optics, 
  Wigner Research Centre for Physics, 
  H-1121 Budapest, 
  Hungary}

\begin{document}

\title{Magnetic properties of hematite revealed by an \fremdwort{ab initio} parameterized spin model} 

\author{Tobias Dannegger} 
\email{tobias.dannegger@uni-konstanz.de} 
\affiliation{\affilUKN} 

\author{András Deák} 
\affiliation{\affilBME} 

\author{Levente Rózsa} 
\altaffiliation[Present address: ]{\affilWRC.} 
\affiliation{\affilUKN} 

\author{E. Galindez-Ruales} 
\affiliation{\affilJGU} 

\author{Shubhankar Das} 
\affiliation{\affilJGU} 

\author{Eunchong Baek} 
\affiliation{\affilJGU} 

\author{Mathias Kläui} 
\affiliation{\affilJGU} 

\author{László Szunyogh} 
\affiliation{\affilBME} 
\affiliation{\affilMTA} 

\author{Ulrich Nowak} 
\affiliation{\affilUKN} 

\date{\today} 

\begin{abstract} 
Hematite is a canted antiferromagnetic insulator, promising for applications in spintronics. 
Here, we present \fremdwort{ab initio} calculations of the tensorial exchange interactions of hematite and use them to understand its magnetic properties by parameterizing a semiclassical Heisenberg spin model. 
Using atomistic spin dynamics simulations, we calculate the equilibrium properties and phase transitions of hematite, most notably the Morin transition. 
The computed isotropic and Dzyaloshinskii--Moriya interactions result in a Néel temperature and weak ferromagnetic canting angle that are in good agreement with experimental measurements. 
Our simulations show how dipole-dipole interactions act in a delicate balance with first and higher-order on-site anisotropies to determine the material's magnetic phase. 
Comparison with spin-Hall magnetoresistance measurements on a hematite single-crystal reveals deviations of the critical behavior at low temperatures. 
Based on a mean-field model, we argue that these differences result from the quantum nature of the fluctuations that drive the phase transitions. 
\end{abstract} 

\maketitle

\section{Introduction} 

As a prototypical weak ferromagnet, the insulating iron oxide hematite ($\alpha$-Fe$_2$O$_3$), one of the main components of rust and the most common iron ore, has interested physicists for a long time. 
Despite its magnetic order being essentially antiferromagnetic, it was shown by Morin~\cite{Morin1950} that a small net magnetic moment emerges above a critical temperature $T_\text{M} \approx \SI{250}{\kelvin}$. 
A new type of magnetic interaction could later explain this phase transition, the Dzyaloshinskii--Moriya interaction (\acronym{DMI})~\cite{Dzyaloshinskii1958, Moriya1960b}, which induces a small canting between the magnetic sublattices. 
This canted antiferromagnetic state is known as the weak ferromagnetic phase. 
In contemporary research on antiferromagnetic spintronics, hematite has shown a remarkable propagation length of magnetic spin currents~\cite{Lebrun2018, Lebrun2020}, among many exciting properties~\cite{Wittmann2022Preprint}. 

The main purpose of this work is to provide a microscopic spin model for this important material. 
While earlier work exists that estimates Heisenberg interaction parameters both experimentally~\cite{Samuelsen1970} and theoretically~\cite{Mazurenko2005,  Logemann2017}, our goal is both to provide a full and detailed parameterization for an atomistic spin model and to validate that model against measurements by simulating critical phenomena but also to shed light on the microscopic origin of those phase transitions. 

Our work begins by calculating tensorial Heisenberg interactions for 170 neighbors (up to the 34th coordination sphere) as well as the spin and orbital magnetic moment for each iron atom in the unit cell \fremdwort{ab initio}. 
Dipole-dipole interactions can then be computed from the crystal structure and the \fremdwort{ab initio} calculated magnetic moments. 
For the on-site anisotropy parameters, the accuracy of the \fremdwort{ab initio} calculations has proved insufficient. 
We solve this issue by fitting our model to angle-dependent measurements of the spin-flop fields instead. 

The remainder of this work is structured as follows: 
After introducing the crystal structure and magnetic properties of hematite, we begin by outlining the experimental methodology of our spin-Hall magnetoresistance measurements in Sec.~\ref{sec:Experiments}. 
Sec.~\ref{sec:AbInitio} then describes the \fremdwort{ab initio} calculations and their results. 
In Sec.~\ref{sec:SpinDynamics}, we apply these results in atomistic spin dynamics simulations to discuss the equilibrium properties of hematite and the origins of its phase transitions. 
We then compare these results to measurements. 
Finally, in Sec.~\ref{sec:QuantumEffects}, we discuss how the quantum nature of the thermal fluctuations leads to a critical behavior in the low temperature regime that is measurably different from a conventional classical prediction.

\begin{figure} 
\begin{center} 
\includegraphics[scale=1]{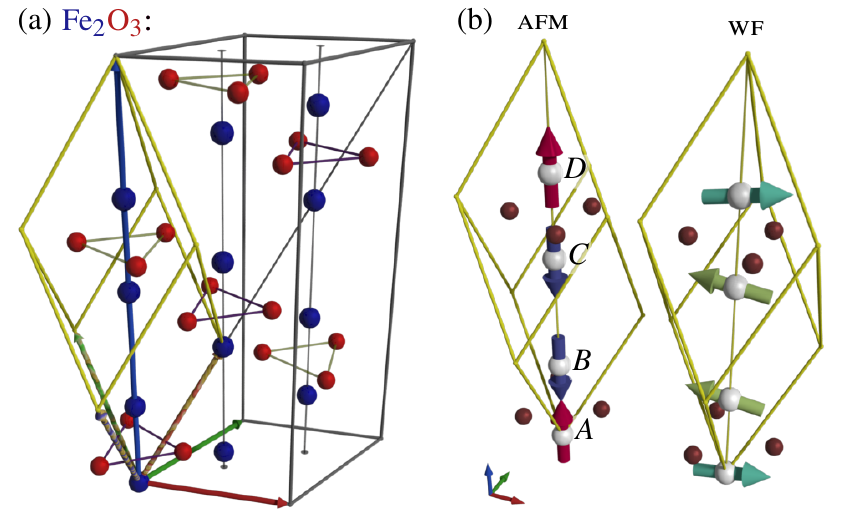} 
\end{center} 
\caption{
(a) The crystal structure of hematite with the conventional hexagonal (gray) and the primitive rhombohedral (yellow) unit cell. 
The primitive basis consists of four iron atoms (blue) and six oxygen atoms (red). 
(b) Orientation of the spin vectors in the \acronym{afm} (left) and \acronym{wf} (right) state. The four Fe sublattices in the primitive rhombohedral unit cell are labeled by $A$, $B$, $C$ and $D$. 
} 
\label{fig:CrystalStructure} 
\end{figure}

Hematite ($\alpha$-Fe$_2$O$_3$) crystallizes in the corundum structure (space group 167, R$\bar{3}$c), which belongs to the hexagonal crystal family. 
\FFig~\ref{fig:CrystalStructure}a visualizes the structure within the hexagonal unit cell, whose $c$ axis is the crystal's highest symmetry axis. 
It also shows the primitive rhombohedral unit cell, whose diagonal lies along this symmetry axis. 
The oxygen atoms mediate the exchange interaction between the iron atoms but do not carry permanent magnetic moments themselves. 
Therefore we do not treat them explicitly in the spin model. 
The magnetic iron atoms are lined up along the $c$ axis and form four magnetic sublattices (labeled $A$ to $D$). 
In the antiferromagnetic (\acronym{afm}) ground state, illustrated in \fig~\ref{fig:CrystalStructure}b, the spins of all four Fe atoms are aligned collinearly along the $c$ axis, with $A$ and $D$ being antiparallel to $B$ and $C$. 
The resulting magnetic structure can be described as double layers of ferromagnetic alignment parallel to the $c$ plane, stacked antiferromagnetically along the $c$ direction. 
In the weak ferromagnetic (\acronym{wf}) phase, the magnetic moments reorient into the basal plane and the \acronym{DMI} induces a small canting between the antiparallel sublattices. The resulting weak magnetization lies in the basal plane too, unless an external magnetic field with an out-of-plane component is applied.

\section{Measurements} \label{sec:Experiments} 

The spin-Hall magnetoresistance (\acronym{SMR}) technique can probe the magnetic state of bilayer systems consisting of hematite (antiferromagnet) and platinum (heavy metal). 
The spin-flop field is detectable using this method as a first-order transition~\cite{Lebrun2019}. 
The single crystal of hematite was obtained commercially with an R-cut orientation (i.e., a \SI{33}{\degree} tilting between the crystallographic $c$ axis and the surface plane). 
The used Hall bars were patterned perpendicular to the projection of the Néel vector using electron beam lithography. 
A subsequent deposition and lift-off of \SI{7}{\nano\metre} platinum were followed by a contacting procedure using a bilayer of chromium (\SI{6}{\nano\metre}) and gold (\SI{32}{\nano\metre}). 
The sample was coupled to a piezo-rotating element in a cryostat with a superconducting magnet capable of variable fields up to \SI{17}{\tesla} and cooled with liquid helium. 
The temperature stability during the measurements reached maximum variations of $\pm \SI{0.05}{K}$ measured with a Cernox sensor element, and the \acronym{SMR} magnitude is in the previously reported order of $10^{-4}$ \cite{Ross2020}.

\section{\fremdwort{Ab initio} calculations} \label{sec:AbInitio} 

\subsection{Self-consistent calculations} 

We performed first-principles calculations for hematite in terms of the screened Korringa--Kohn--Rostoker (\acronym{SKKR}) multiple scattering theory~\cite{ZabloudilKKRBook} in the atomic-sphere approximation (\acronym{ASA}). 
The bulk crystal structure is assembled using the conventional hexagonal unit cell, see \fig~\ref{fig:CrystalStructure}a. 
The hexagonal lattice parameters $a = \SI{5.067}{\angstrom}$ and $c = \SI{13.882}{\angstrom}$ were chosen to match the structure optimized by Rohrbach \fremdwort{et al.}\ using $\text{\acronym{GGA}}+U$~\cite{Rohrbach2004}. 
According to the suggestion of Sandratskii \fremdwort{et al.}~\cite{Sandratskii1996a}, in order to achieve sufficient space filling within the \acronym{ASA} we added \enquote{empty} atomic spheres (\acronym{ES}) between the Fe atoms labeled by $B$ and $C$, as well as between $A$ and $D$. 
The hexagonal unit cell in our calculations contained thus 36 atomic spheres (12 Fe, 6 \acronym{ES} and 18 O). 

We carried out self-consistent field (\acronym{scf}) calculations for the ordered \acronym{afm} state of hematite with the magnetic orientation pointing along the $c$ axis, as well as for the paramagnetic state by employing the Disordered Local Moment (\acronym{DLM}) theory~\cite{Gyorffy1985, Staunton2006}. 
We then used the Spin-Cluster Expansion (\acronym{SCE}) to extract spin model parameters from the \acronym{DLM} state by mapping the adiabatic energy surface of the fluctuating state onto a Heisenberg model~\cite{Drautz2004, Szunyogh2011}. This method has been used successfully to describe \acronym{afm}-\acronym{fm} interfaces such as exchange bias systems~\cite{Szunyogh2011, Simon2018}, as well as bulk noncollinear antiferromagnets~\cite{Nyari2019, Simon2020}.

For the partial waves within the multiple scattering theory, we used an angular momentum cutoff of $\ell_\text{max}=2$. 
The effective potentials and fields were constructed within the generalized gradient approximation (\acronym{GGA}) as parameterized according to Perdew, Burke, and Ernzerhof~\cite{Perdew1996}. 
To account for the strong Coulomb repulsion of the Fe $d$ electrons, we employed the Hartree--Fock approximation ($\text{\acronym{GGA}}+U$) with the parameters $U = \SI{6}{\electronvolt}$ and $J = \SI{2}{\electronvolt}$. 
Note that the value  $U - J = \SI{4}{\electronvolt}$ is commonly accepted in the literature~\cite{Bandyopadhyay2004, Rollmann2004, Rohrbach2004, Mazurenko2005}. 
With this choice, we obtained a band gap of \SI{2.29}{\electronvolt} in excellent agreement with experimental values of \SIrange{2.14}{2.2}{\electronvolt}~\cite{Gilbert2009}.  
The necessary energy integrations were performed by sampling 16 points along a semicircular contour in the upper complex semiplane. 
During the self-consistent iterations at every energy point, we used  9450 $k$ points for the integrations in the hexagonal Brillouin zone, whereas for the calculations of the spin model parameters and the magnetic anisotropy, we gradually increased the number of $k$ points  up to about \num{440000} near the Fermi energy.

In our self-consistent calculations, the Fermi energy $E_\text{F}$ was underestimated by about \SI{0.5}{\electronvolt} compared to the bottom of the insulating gap. 
This is a well-known shortcoming of \acronym{KKR} Green's function calculations due to the insufficient angular momentum convergence in the evaluation of the charge density. 
Unfortunately, using an angular momentum cutoff higher than $\ell_\text{max} = 2$ was not possible, as the combination of a fully relativistic description and the very large unit cell led to a memory demand we could not increase further in our \acronym{SKKR} code. 
Zeller proposed a procedure to rescale the energy-dependent contributions of the charge density by validating the total charge using Lloyd's formula~\cite{Zeller2007}. 
However, for similar reasons as above, this approach is computationally not feasible for our \acronym{SKKR} implementation. 
As compared to the width of the valence band of about \SI{7.5}{\electronvolt}, the error of the calculated  Fermi level amounts  to approximately \SI{7}{\percent}. 

In order to mimic the insulating state of hematite, we simply set $E_\text{F}$ to the bottom of the band gap by keeping the self-consistently calculated effective potentials and fields fixed. 
The validity of this choice of $E_\text{F}$ is also supported by the fact that the spin model obtained using the self-consistent Fermi level ($E^\text{scf}_\text{F}$) turned out to have a ferromagnetic ground state, whereas the spin model derived by using the corrected Fermi level ($E^\text{corr}_\text{F}$) provided the correct \acronym{afm} ground state as sketched in \fig~\ref{fig:CrystalStructure}b. 

\subsection{Atomic magnetic moments} 

\begin{table} 
  \caption{
  Spin ($m_\text{spin}$) and orbital ($m_\text{orb}$) magnetic moments of the Fe and O atoms from the \acronym{SKKR} $\text{\acronym{GGA}}+U$ calculations in the \acronym{afm} and \acronym{DLM} state using the \acronym{scf} and the corrected Fermi level (see text). The results of earlier \fremdwort{ab initio} calculations are also shown for comparison. The value shared between an Fe and an O column refer to the spin magnetic moment per Fe atom for the given method. Experimental values refer to the total magnetic moment per Fe atom. All values are given in $\mu_\text{B}$. 
  } 
  \begin{tabular}{ l  c c c c } 
\toprule 
Source  & \quad $m^\text{Fe}_\text{spin}$ \qquad  & $m^\text{O}_\text{spin}$  & $m^\text{Fe}_\text{orb}$  & $m^\text{O}_\text{orb}$  \\  
        &                                         & $ (\times 10^{-3}) $      & $(\times 10^{-2})$        & $(\times 10^{-3})$  \\ \toprule 
    \acronym{AFM} ($E^\text{scf}_\text{F}$)               & $4.04$  & $7.08$  & $0.49$  & $0.31$  \\ 
    \acronym{AFM} ($E^\text{corr}_\text{F}$)              & $4.17$  & $0.41$  & $1.13 $ & $0.40 $ \\ 
    \acronym{DLM} ($E^\text{scf}_\text{F}$)               & $4.07$  & $0.00$  & $0.39$  & $0.00$  \\ 
    \acronym{DLM} ($E^\text{corr}_\text{F}$)              & $4.23$  & $0.00$  & $1.06$  & $0.00$  \\ \midrule 
    \acronym{AFM} \acronym{LSDA}~\cite{Sandratskii1996a}  & $3.69$  &         & $3 $    &         \\  
    \acronym{AFM} $\text{\acronym{LSDA}}+U$~\cite{Mazurenko2005} & \multicolumn{2}{c}{$4.1$}  & & \\ 
    \acronym{AFM} $\text{\acronym{GGA}}+U$~\cite{Logemann2017}    & $4.09$  & $0$             & & \\ \midrule 
    experiment~\cite{Kren1965, Coey1971}  & \multicolumn{4}{c}{$4.6-4.9$}   \\ 
        experiment~\cite{Hill2008}        &  \multicolumn{4}{c}{$4.22$}     \\ 
  \bottomrule 
  \end{tabular} 
  \label{tab:ab_initio_moments} 
\end{table} 
The spin and orbital moments we obtained in terms of the \acronym{SKKR} $\text{\acronym{GGA}}+U$ method are shown in the first four rows of \tab~\ref{tab:ab_initio_moments}. 
Clearly, there is only a minor difference in the Fe spin moments between the ordered \acronym{afm} and the \acronym{DLM} states, which may be attributed to the nearly occupied majority spin band of Fe in both cases. 
The rows labeled by $E^\text{scf}_\text{F}$ and $E^\text{corr}_\text{F}$ correspond to calculations with the self-consistently calculated (\enquote{incorrect}) Fermi energy and to those where the Fermi level was shifted to the bottom of the gap, respectively, as explained above. 
The Fe spin moments calculated with  $E^\text{scf}_\text{F}$ are in agreement with earlier $\text{\acronym{LSDA}}+U$, or $\text{\acronym{GGA}}+U$ calculations~\cite{Mazurenko2005, Logemann2017}. 
They are further increased by about \num{0.15} $\mu_\text{B}$ when shifting the Fermi level to the band bottom, bringing the result closer to the experimental values~\cite{Kren1965, Coey1971, Hill2008}. 
The underestimation of the magnetic moments seen in experiments is a common feature of existing theoretical works in the literature. 
But compared to more recent measurements by Hill \fremdwort{et al.}~\cite{Hill2008}, our magnetic moment seems in excellent agreement with experiments. 
Regarding orbital moments, we find that they are at least two orders of magnitude smaller than spin moments. 

Even with the incorrect value of the Fermi level, our calculated spin moments come considerably closer to measured values than in early \acronym{LSDA} calculations~\cite{Sandratskii1996a}, demonstrating the need for incorporating electron correlations in order to describe the magnetism of hematite correctly.

\subsection{Ground state and weak ferromagnetism} \label{sec:AbIniWeakFerromagnetism}

By fixing the \acronym{scf} effective potentials and fields in the \acronym{afm} configuration with magnetization parallel to the $c$ axis, in the spirit of the magnetic force theorem, we calculated the band energy by changing the angle $\theta$ of the magnetization relative to the $c$ axis, 
\begin{equation}  
E_\text{band}(\theta) = \int_{\epsilon_\text{bott}}^{E^\text{corr}_\text{F}} \, \text{d}\epsilon (\epsilon - E^\text{corr}_\text{F}) \, n(\epsilon, \theta) \, 
\end{equation}  
where $n(\epsilon,\theta)$ is the density of states (\acronym{DOS}) and $\epsilon_\text{bott}$ is chosen below the bottom of the valence band. 
According to the trigonal symmetry of the lattice, our calculations showed an angle dependence $E_\text{band}(\theta) = - d_2 \cos^2(\theta)$ with high accuracy. 
We obtained a value for the uniaxial magnetocrystalline anisotropy (\acronym{MCA}) energy of $d_2 = E_\text{band}(90^\circ) - E_\text{band}(0^\circ) = \SI{40.49}{\micro\electronvolt}$ per Fe atom favoring a magnetization parallel to the $c$ axis. 
Thus, our calculations of the \acronym{MCA} predict an out-of-plane \acronym{afm} order as the ground state of hematite, matching experimental findings.

Orienting the Fe moments in the plane allows us to further decrease the energy of the \acronym{afm} configuration by canting the moments of the two Fe \acronym{afm} sublattices into the perpendicular in-plane direction, forming a \acronym{wf} state. 
\begin{figure} 
\vspace*{1em} 
\begin{center} 
\includegraphics[scale=1]{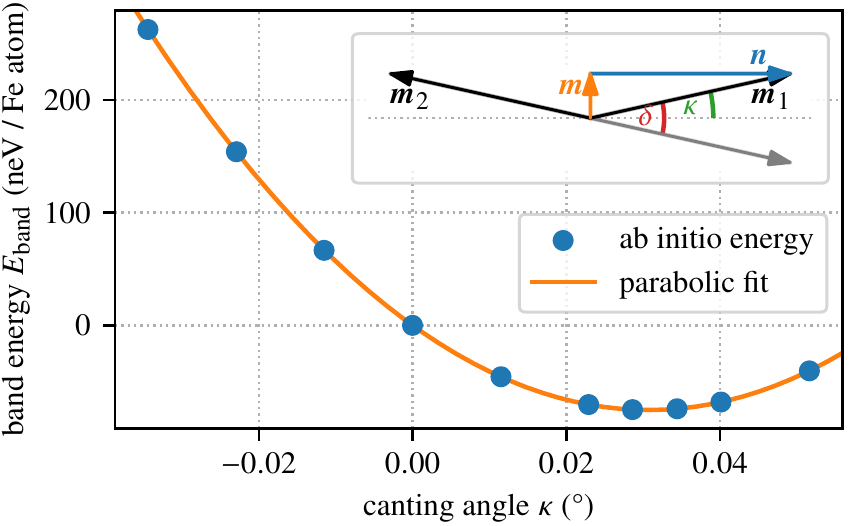} 
\end{center} 
\vskip -12pt 
\caption{
Energy as a function of weak ferromagnetic distortion angle $\kappa$ as obtained from magnetic force theorem calculations. 
Inset: magnetization vectors $\vec{m}_1$ and $\vec{m}_2$ of the magnetic double layers, net magnetization $\vec{m}$ and Néel vector $\vec{n}$. The canting angle with respect to the collinear configuration is denoted by $\kappa$. 
} 
\label{fig:canting_angle} 
\label{fig:energy_WF} 
\end{figure} 
By varying the canting angle $\kappa$ (cf.\ \fig~\ref{fig:canting_angle}) we indeed obtain an energy minimum at $\kappa=\SI{0.031}{\degree}$ 
(or \SI{0.54}{\milli\radian}), in excellent agreement with earlier theoretical findings~\cite{Sandratskii1996b, Mazurenko2005}. 
The energy difference between the canted \acronym{wf} state and the collinear \acronym{afm} state is only \SI{74.7}{\nano\electronvolt} per Fe atom, about three orders of magnitude smaller than the uniaxial anisotropy. 
These energy scales underpin the picture that the Morin transition is primarily a reorientation transition from the out-of-plane \acronym{afm} order to an in-plane orientation driven by the different temperature dependence of various contributions to the anisotropy, and once the system is in the in-plane state, the canting is induced by the \acronym{DMI}.

\subsection{Exchange tensors} \label{sec:ExchangeTensors} 

The \acronym{SCE} based on the relativistic \acronym{DLM} scheme~\cite{Szunyogh2011} provides us with a bilinear tensorial Heisenberg model of the form 
\begin{equation} 
  \hamil 
  = -\frac{1}{2} \sum_{i \neq j} \vec{S}_i^T \tensorial{J}_{ij} \vec{S}_j 
  - \sum_i \vec{S}_i^T \tensorial{K}_i \vec{S}_i \, , 
\end{equation} 
where $\tensorial{J}_{ij}$ is the exchange interaction tensor and $\tensorial{K}_i$ is the on-site anisotropy matrix.   
The interaction term can be decomposed into three parts according to the spherical tensor components of $\tensorial{J}_{ij}$, namely 
\begin{align} 
 & \tensorial{J}_{ij} = J_{ij}^\text{iso} I_3 + \tensorial{J}_{ij}^\text{A} + \tensorial{J}_{ij}^\text{S} \nonumber\\ 
         &= \tensorial{I}_3 \tfrac{1}{3}\tr \tensorial{J}_{ij} + \tfrac{1}{2}\left(\tensorial{J}_{ij} - \tensorial{J}_{ij}^T\right) + \tfrac{1}{2}\left(\tensorial{J}_{ij} + \tensorial{J}_{ij}^T - \tfrac{2}{3} \tensorial{I}_3 \tr \tensorial{J}_{ij}\right) \, , 
\end{align} 
where the first term is the isotropic part with the identity matrix $\tensorial{I}_3$, the second term is the antisymmetric part, and the last term is the traceless symmetric part of the exchange tensor. 
These terms correspond to the isotropic Heisenberg interaction, the Dzyaloshinskii--Moriya (\acronym{DM}) interaction~\cite{Dzyaloshinskii1958, Moriya1960b} and the two-ion anisotropy, respectively. 
In particular, the \acronym{DM} vectors can be defined as the vector invariant of the exchange tensors, 
\begin{align} 
  \vec{D}_{ij} = \left(J_{ij, zy}^\text{A}, J_{ij,xz}^\text{A}, J_{ij,yx}^\text{A}\right), 
\end{align} 
corresponding to the energy term $\vec{D}_{ij} \cdot \left(\vec{S}_i \times \vec{S}_j\right)$. 
In line with the uniaxial \acronym{MCA} energy, $d_2$, the site-dependent uniaxial two-ion anisotropy energy can be defined as 
\begin{align} 
\Delta E_{\text{tia},ij}= - \sigma_{ij} (J_{ij,zz}- J_{ij,xx}) \, , 
\end{align} 
where $\sigma_{ij}$ denotes the sign resulting from the relative orientation of the interaction partners ($+1$ for parallel, $-1$ for antiparallel spins). 

There are certain symmetry constraints that the exchange tensors should fulfill: they should be invariant under symmetry operations from the crystal's space group and the \acronym{DMI} component should satisfy Moriya's five symmetry rules~\cite{Moriya1960b}. 
Reassuringly, the results from our \fremdwort{ab initio} calculations possess all these symmetries but there are tiny inaccuracies in the \si{\nano\electronvolt} order. 
These deviations, albeit small, can lead to artifacts in the later spin dynamics simulations, such as lifting the degeneracy between symmetrically equivalent states, or a ground state that is ever so slightly tilted to the crystal axis (by \SI{0.37}{\degree}). 
To avoid these issues, we symmetrized the exchange tensors by enforcing Moriya's symmetry rules and taking the mean of all symmetric equivalents for each interaction pair. 

\begin{figure} 
\begin{center} 
\includegraphics[scale=1]{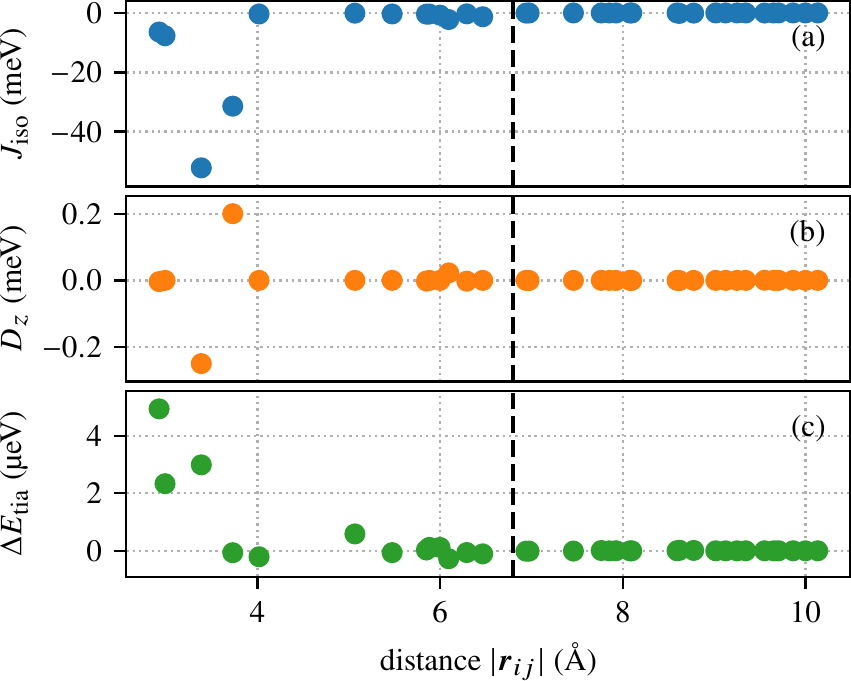} 
\end{center} 
\caption{
\fremdwort{Ab initio} calculated interaction energy per atom, split into (a) the isotropic exchange energy, (b) the out-of-plane component of the \acronym{DM} vector, and (c) the two-ion anisotropy expressed as the energy difference between out-of-plane and in-plane alignment of the magnetic moments. 
The dashed vertical line marks the cutoff radius of \SI{6.8}{\angstrom} used for the spin dynamics simulations. 
} 
\label{fig:jij_vs_distance} 
\end{figure}

The spatial distribution of the Fe-Fe interactions is shown in \fig~\ref{fig:jij_vs_distance}. 
The isotropic couplings are about a hundred times larger than the magnitudes of the \acronym{DM} vectors, and the two-ion anisotropy is another order of magnitude smaller. 
Among the abundance of isotropic exchange interactions the dominant ones are the third and fourth nearest neighbor shells, which provide strong \acronym{afm} couplings between Fe atoms on opposite magnetic sublattices, robustly preferring the \acronym{afm} order seen in experiments. 
A mean-field estimate based on the Fourier transform of the exchange tensors predicts the same \acronym{afm} order with a mean-field Néel temperature of \SI{1259}{\kelvin}. 

\begin{table} 
  \caption{
Calculated isotropic exchange interactions for the five nearest atomic shells, in \si{\milli\electronvolt} units. The interactions are listed in the same order (in increasing order of interatomic distance) as in \tab~I of Ref.~\cite{Mazurenko2005}. 
  } 
  \begin{tabular}{lcSSS} 
  \toprule 
           & sublattices  & {this work} & {Ref.~\cite{Logemann2017}}  & {Ref.~\cite{Mazurenko2005}} \\ \midrule 
    $J_1$  & $A$-$B$          &  -3.21      &  -3.5                       &  -8.58                      \\ 
    $J_2$  & $A$-$D$          &  -3.84      &  -3.2                       &   7.3                       \\ 
    $J_3$  & $A$-$B$          & -26.10      & -13.9                       & -25.22                      \\ 
    $J_4$  & $A$-$C$          & -15.71      &  -9.8                       & -17.5                       \\ 
    $J_5$  & $A$-$D$          &  -0.17      &                             &   0.07                      \\ \bottomrule 
  \end{tabular}

  \label{tab:isotropic_heisenberg} 
\end{table} 

The first five isotropic couplings are also collected in \tab~\ref{tab:isotropic_heisenberg} for comparison with earlier theoretical results. 
We find a comparatively similar spatial dependence as Logemann \fremdwort{et al.}~\cite{Logemann2017}, but the values of the dominant $J_3$ and $J_4$ interactions are almost twice as large as what they found. 
This is also reflected in the mean-field Néel temperature of \SI{878}{\kelvin} in Ref.~\cite{Logemann2017}, which is even lower than the experimental value. 
In contrast, our dominant interactions are very similar to those found by Mazurenko and Anisimov~\cite{Mazurenko2005}, however, the interactions for the first two shells are less than half of theirs (and both \acronym{afm}) in our calculations.

\begin{figure} 
\begin{center} 
\includegraphics{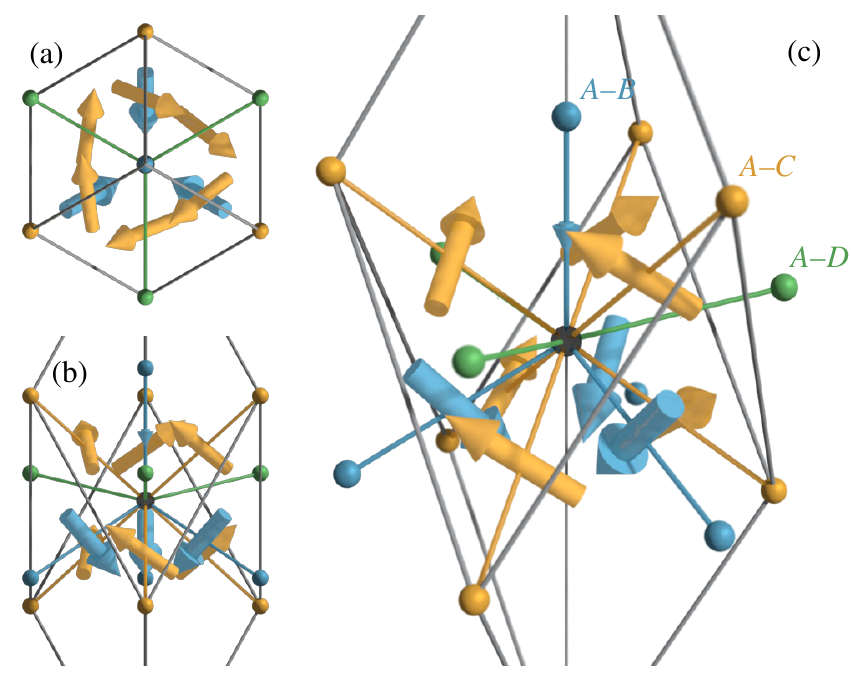} 
\end{center} 
\caption{
Orientation of the \fremdwort{ab initio} computed Dzyaloshinskii--Moriya vectors. 
Shown are all neighbors belonging to the first four shells, relative to an atom of sublattice $A$ (shown in black at the center). 
The \acronym{DMI} between atoms belonging to sublattices $A$ and $D$ is zero because there are inversion centers between the interaction partners. 
Depicted are orthographic projections from the (a) \crystaldir{001} and (b) \crystaldir{100} directions as well as a perspective projection (c). 
} 
\label{fig:DMVectors} 
\end{figure}

The three-dimensional configuration of the \acronym{DM} vectors is rendered in \fig~\ref{fig:DMVectors}. 
The dominant contributions come from the third and fourth shells surrounding the central atom, with three and six sites on the shells, respectively. 
The threefold rotational symmetry of the crystal is nicely reflected in the \acronym{DM} vector configuration implying that the effective \acronym{DM} interaction only arises through the $z$ component of the vectors. 
Since the \acronym{DM} interaction prefers canting of the corresponding spins in a plane perpendicular to the axis of the \acronym{DM} vector, this also explains why the \acronym{wf} distortion only appears in an in-plane state. 
We also note that there is an inversion center between atoms $B$ and $C$ in the rhombohedral unit cell (cf.\ \fig~\ref{fig:CrystalStructure}), implying that the \acronym{DM} interaction between atoms $B-C$ and atoms $A-D$ in the unit cell is exactly zero.

\subsection{Magnetocrystalline anisotropy} \label{sec:AbIniAni} 

As for the anisotropies, $C_3$ symmetry restricts the second-order on-site anisotropy to a uniaxial form, and further considering space group symmetries connecting the four Fe sites in the rhombohedral cell we only have 
\begin{equation} 
  - \sum_i \vec{S}_i^T \tensorial{K}_i \vec{S}_i = - d_2 \sum_i S_{i,z}^2. 
\end{equation} 
The total uniaxial two-site anisotropy is given by 
\begin{equation} 
  \Delta E_\text{tia} =  \sum_{j (\ne i)} \Delta E_{\text{tia},ij} \, ,  
\end{equation} 
and is the same for all sublattices. 

From the \acronym{SCE} calculations we obtain $d_2 = \SI{-2.24}{\micro\electronvolt}$ and $\Delta E_\text{tia} = \SI{-11.20}{\micro\electronvolt}$ normalized to one Fe atom.  
The sum of the second-order anisotropy arising from the on-site and two-ion contributions of the \fremdwort{ab initio} spin model is \SI{-13.44}{\micro\electronvolt}, i.e., it favors an in-plane orientation for the ground state magnetization. 
This contrasts with our magnetic force theorem calculations performed in the ordered \acronym{afm} state (cf.\ Sec.~\ref{sec:AbIniWeakFerromagnetism}), which predicts an easy $c$ axis anisotropy for the ground state in agreement with the experiments. 

This disagreement leads us to conclude that the \acronym{sce} calculations lack the necessary accuracy on the relevant energy scale of \SI{10}{\micro\electronvolt} (relevant for the \acronym{MCA}). 
Furthermore, we know that the description of the transversal spin-flop transition as a first-order phase transition requires the presence of a fourth-order anisotropy term in the Hamiltonian~\cite{MorrishBook}, but in our \fremdwort{ab initio} force theorem calculations, this term has a completely negligible magnitude. 

Instead, we find the following approach more promising. 
We parameterize the spin model with the tensorial interactions as calculated within the \acronym{SCE}, since they provide a Néel temperature and \acronym{wf} canting angle in good agreement with experiments, as we shall see in the following (the energy scales are also much larger here, in the \si{\milli\electronvolt} range). 
The on-site anisotropy parameters $d_2$ and $d_4$ will be treated as adjustable parameters determined by comparison with experimental measurements of the spin-flop transition. 
The dipolar interactions are calculated from the \fremdwort{ab initio} lattice structure and atomic magnetic moments.

\section{Spin dynamics simulations} \label{sec:SpinDynamics}

Our atomistic spin dynamics simulations are based on an extended Heisenberg model of the form 
\begin{equation} 
\hamil
= -\frac{1}{2} \sum_{i \neq j} \vec{S}_i^T \tensorial{J}_{ij} \vec{S}_j 
- d_2 \sum_i S_{i,z}^2 
- d_4 \sum_i S_{i,z}^4 
- \mu_s \vec{B} \cdot \sum_i \vec{S}_i 
. 
\label{eq:Hamiltonian} 
\end{equation} 
Here, the $\tensorial{J}_{ij}$ are exchange tensors that contain the isotropic exchange, \acronym{DMI}, and two-ion anisotropy from the \acronym{SKKR} calculations (cf.\ Sec.~\ref{sec:ExchangeTensors}) as well as dipole-dipole interactions (see the following Sec.~\ref{sec:ddi}). 
The next two terms model the second and fourth order on-site anisotropies as discussed in the previous section. 
The last term is the Zeeman energy from external magnetic fields $\vec{B}$ and uses a magnetic moment per iron atom of $\mu_s = \num{4.2313}\ \mu_\text{B}$, which is the sum of the spin and orbital moments computed from the \acronym{DLM} state with the corrected Fermi level (see \tab~\ref{tab:ab_initio_moments}). 

The spin dynamics are then simulated by integrating the stochastic Landau--Lifshitz--Gilbert (\acronym{LLG}) equation~\cite{LandauLifshitz1935, Gilbert2004, Brown1963, Nowak2007} with a damping parameter of $\alpha = \num{0.001}$. 
This value is larger than what is usually assumed for hematite~\cite{Lebrun2019}, 
but it leads to a faster relaxation toward equilibrium, and so long as we are only concerned with equilibrium states, the choice of $\alpha$ does not affect the results.

\subsection{Dipole-dipole interaction} \label{sec:ddi} 
\begin{figure} 
\begin{center} 
\includegraphics{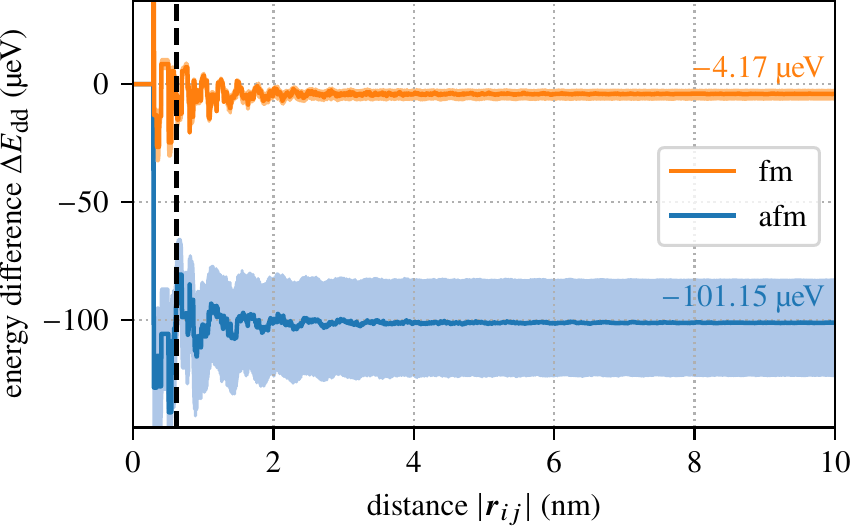} 
\end{center} 
\caption{Effective anisotropy due to dipole-dipole interactions. 
Shown is the energy difference per spin between the out-of-plane and in-plane state for both antiferromagnetic and ferromagnetic alignment of the sublattices, in dependence of the interaction range $|\vec{r}_{ij}|$. 
The total bulk values given in the graph have been calculated from the cumulative dipole-dipole energies up to a distance of \SI{1}{\micro\metre}. 
The negative sign of both values indicates that the dipole-dipole interaction favors the in-plane orientation of both the Néel vector and the magnetization. 
The dashed vertical line marks the cutoff radius of \SI{6.2}{\angstrom} used for the spin dynamics simulations. 
The shaded areas indicate the uncertainty interval resulting from a \SI{1}{\percent} relative uncertainty of the lattice parameters and the atomic magnetic moment. 
} 
\label{fig:DipoleDistance} 
\end{figure}

The energy contribution from dipole-dipole interactions between the atomic magnetic moments has the form 
\begin{equation} 
\hamil_\text{ddi} = - \frac{\mu_s^2 \mu_0}{8 \pi} \sum_{i \neq j} \frac{3 (\vec{S}_i \cdot \vec{r}_{ij}) (\vec{r}_{ij} \cdot \vec{S}_j)}{\abs{\vec{r}_{ij}}^5} - \frac{\vec{S}_i \cdot \vec{S}_j}{\abs{\vec{r}_{ij}}^3} 
, 
\end{equation} 
where $\vec{r}_{ij}$ is the distance vector between two lattice sites $i$ and $j$, and $\mu_0$ is the vacuum permeability. 
The effect of this interaction is an energy difference, i.e., an effective two-site anisotropy, between the out-of-plane and in-plane orientation of the magnetic moments. 
\FFig~\ref{fig:DipoleDistance} shows the energy difference between the in-plane and the out-of-plane orientation of the Néel and magnetization vector as a function of the cutoff radius. 
The sum of all interactions within a sphere of \SI{1}{\micro\metre} radius amounts to \SI{-101.15}{\micro\electronvolt} and \SI{-4.17}{\micro\electronvolt}, respectively. 
These rather large values are the result of the magnetic structure of hematite consisting of double layers of ferromagnetic alignment. 
In each of these layers, the magnetic moments can minimize their energy by assuming a nose-to-tail rather than a broadside configuration. 
Hence, the dipole-dipole interaction leads to a preference for the in-plane state. 

The effective anisotropy we calculated from the dipolar interactions is about \SI{30}{\percent} smaller than an earlier calculation by Artman, Murphy, and Foner~\cite{Artman1965}. 
Most of this deviation comes from the assumed magnetic moment per Fe atom, which in our case is \SI{10}{\percent} smaller than for Artman \fremdwort{et al}. 
Their calculations were also based on a different set of lattice parameters~\cite{NewnhamDeHaan1962} with lattice constants that are slightly smaller (by less than \SI{1}{\percent}). 

For the spin dynamics simulations, dipole-dipole interactions were taken into account up to a range of \SI{6.2}{\angstrom}, in order to reduce the computational effort. 
The resulting deviation from the total dipole-dipole energy is below \SI{8}{\percent}. 
This deviation is acceptable when compared to the uncertainty of both the dipole-dipole energies and the two-site anisotropy that results from the anisotropic part of the \acronym{SCE} exchange tensors.

\subsection{Anisotropy} \label{sec:Anisotropy} 

We have seen that the two-ion anisotropies (dipole-dipole interaction and symmetric anisotropic exchange) energetically favor an in-plane alignment of the magnetic moments. 
These must be compensated by larger, positive on-site anisotropies in the ground state, which has the magnetic moments aligned out of plane. 
As discussed in Sec.~\ref{sec:AbIniAni}, we have the second- and fourth-order anisotropy energies, $d_2$ and $d_4$, left as free parameters. 
There is also a sixth-order triaxial basal plane anisotropy~\cite{Banerjee1963, FlandersSchuele1964}, which we neglect because it is very small (around \SI{1}{\nano\electronvolt}) and it does not qualitatively alter any of the phase transitions. 

To determine $d_2$ and $d_4$, we look at \acronym{SMR} measurements of the spin-flop transition. 
The spin-flop field $B_\text{sf}$ depends both on the angle $\theta$ between the applied magnetic field and the crystal's $c$ axis and on the temperature $T$, becoming zero at the Morin temperature $T_\text{M}$. 

By minimizing the Hamiltonian as defined in \eq~\eqref{eq:Hamiltonian}, one finds that the longitudinal spin-flop field ($\theta = \SI{0}{\degree}$) depends mostly on the sum of the anisotropy energies, $d_2 + d_4$, while the transversal spin-flop field ($\theta = \SI{90}{\degree}$) is very susceptible to the contribution of $d_4$. 
Therefore, angle-dependent spin-flop measurements are ideally suited to determine these parameters. 

Our approach is hence to first adjust our model's anisotropy parameters to angle-dependent measurements of the spin-flop field $B_\text{sf}(\theta)$ at temperatures closely below the Morin temperature $T_\text{M}$. 
Details on how the spin-flop transition is simulated can be found in App.~\ref{app:SpinFlopSims}. 
Having fixed the free parameters, we can then evaluate the model's critical behavior across the whole temperature range $(0, T_\text{M})$ and compare it to temperature-dependent measurements of $B_\text{sf}(T)$. 
\begin{figure} 
\begin{center} 
\includegraphics{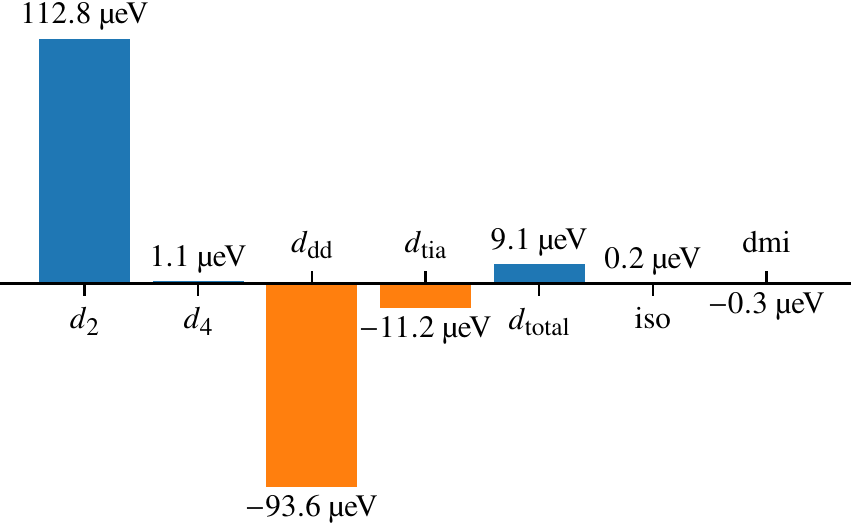} 
\end{center} 
\caption{Contributions to the internal energy difference between the \acronym{wf} and \acronym{afm} state (without external fields). 
Both the dipole-dipole interaction $d_\text{dd}$ and the two-ion anisotropy $d_\text{tia}$ favor an in-plane orientation of the magnetic moments while the second- and fourth-order on-site anisotropies lead to a preference of the out-of-plane state in total. 
The isotropic exchange and \acronym{DMI} energy are also affected by the transition from the \acronym{afm} to the \acronym{wf} state but their effect is much smaller than that of the anisotropy terms. 
} 
\label{fig:AnisotropyComparison} 
\end{figure} 

The parameters we found are $d_2 = \SI{112.8}{\micro\electronvolt}$ and $d_4 = \SI{1.1}{\micro\electronvolt}$. 
In \fig~\ref{fig:AnisotropyComparison}, we compare these values to the other terms in the Hamiltonian. 
In total, the energy difference between the \acronym{afm} and \acronym{wf} state for $T = 0$, $B = 0$ is only \SI{9}{\micro\electronvolt}, 
which is the sum of several competing terms, by far the largest of which are the second-order on-site anisotropy and the dipole-dipole interaction. 
The competition between these two is what determines the equilibrium state of the material. 
The different temperature dependence of the free energy associated with each term leads to the Morin transition. 
The dipole-dipole energy is the only contribution that has the same order of magnitude as the second-order on-site anisotropy. 
Without it, the system would not exhibit a Morin transition. 

Compared to earlier estimates based on antiferromagnetic resonance measurements~\cite{Morrison1973}, our fourth-order anisotropy energy is about \SI{25}{\percent} smaller, while the total second-order anisotropy (the sum of the quadratic and bilinear terms) is about \SI{50}{\percent} larger. 
Our total anisotropy energy is hence approximately \SI{30}{\percent} lower. 
We cannot expect a better agreement with those earlier estimates, since they are based on effective parameters derived from the comparison of experimental results with mean-field models. 
E.g., the calculation used by Morrison \fremdwort{et al.} is based on an effective exchange field that is \SI{35}{\percent} lower than our \fremdwort{ab initio} result. 
While the resulting Néel temperature seems close to the expected value in the mean-field approximation, it would lead to an underestimation by a third with our atomistic spin dynamics simulations. 
This underlines the advantages of an \fremdwort{ab initio} approach, which greatly reduces the number of free parameters in the model and hence improves the estimation of the remaining parameter values.

\subsection{Néel transition} 

\begin{figure} 
\begin{center} 
\includegraphics{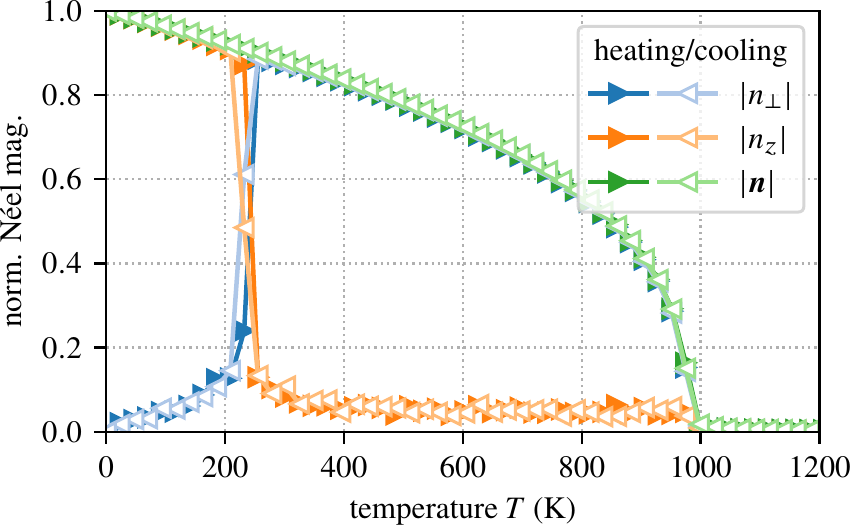} 
\end{center} 
\caption{Magnetization curve of the hematite spin model. 
The plot shows data from two simulations: heating up from the ground state (solid triangles pointing right) and cooling down from the paramagnetic state (white triangles pointing left). 
The Néel vector length $\abs{\vec{n}}$ displays the Néel transition at $T_\text{N} = \SI{989}{\kelvin}$. 
The in-plane component $\abs{n_\perp}$ and the out-of-plane component $\abs{n_z}$ of the Néel vector reveal the Morin transition at a temperature of $T_\text{M} = \SI{240(12)}{\kelvin}$. 
} 
\label{fig:MorinHysteresis} 
\end{figure}

To study the temperature-dependent phase transitions in our model, the system is initialized in the ground state at $T = \SI{0}{\kelvin}$ and then heated in steps of \SI{23.2}{\kelvin}. 
In another simulation, the system is initialized in a paramagnetic state above the Néel temperature nd then cooled down to \SI{0}{\kelvin}. 
At each temperature step, the system is given \SI{50}{\pico\second} to equilibrate and, when equilibrium is reached, the relevant order parameters are averaged over another \SI{50}{\pico\second}. 
The results are given in \fig~\ref{fig:MorinHysteresis}. 

The norm of the Néel vector, $|\vec{n}|$, shows a clear Néel transition (from the weak ferromagnetic to the paramagnetic state), at a temperature of $T_\text{N} = \SI{989}{\kelvin}$. 
This is in good agreement with literature 
values that range from \SIrange{950}{970}{\kelvin}~\cite{MorrishBook}. 
As the Néel temperature is mainly determined by the total isotropic exchange interaction in the system (other contributions, like the anisotropy constant, are several orders of magnitude lower and therefore negligible), we can be confident that this part of the \fremdwort{ab initio} calculation is indeed very accurate.

\subsection{Weak ferromagnetic canting angle} 

While the Néel temperature is a good indicator for the correctness of the isotropic part of the exchange interaction, the antisymmetric part, i.e., the \acronym{DMI}, is reflected in the weak ferromagnetic canting angle $\kappa$. 
Based on the \fremdwort{ab initio} calculated isotropic exchange and \acronym{DM} energies, we can calculate the canting angle within our spin model to be 
\begin{equation} 
\kappa = \frac{1}{2} \arctan\left(\frac{D_\text{eff}}{J_\text{eff}}\right) 
. 
\label{eq:CantingAngle} 
\end{equation} 
Here, the effective \acronym{DM} energy $D_\text{eff}$ is defined as the sum of the $xy$ components of all exchange tensors with respect to a given lattice site and the effective isotropic exchange energy $J_\text{eff}$ is the sum of the $xx$ components of all exchange tensors between lattice sites of opposite spin alignment. 
Therefore, the normalized magnetization in the \acronym{wf} state (without external field) is given by 
\begin{equation} 
m = \sin(\kappa) = \pm \sqrt{\frac{1}{2} \left( 1 - \sqrt{\frac{J_\text{eff}^2}{J_\text{eff}^2 + D_\text{eff}^2}} \right)} 
. 
\label{eq:WeakMagnetisation} 
\end{equation} 

The value produced by the spin model ($\kappa=\SI{0.038}{\degree}$) is slightly larger than the \fremdwort{ab initio} result ($\kappa=\SI{0.031}{\degree}$). 
It is in good agreement with earlier theoretical findings~\cite{Sandratskii1996b, Mazurenko2005}. 
Experimental measurements, however, have previously reported somewhat larger values ($\kappa=\SI{0.0554(8)}{\degree}$~\cite{Hill2008}). 

This shows that the \fremdwort{ab initio} calculated \acronym{DMI} values are also, at least in total, quantitatively accurate, since, as \eq~\eqref{eq:CantingAngle} shows, the canting angle in the \acronym{wf} phase is entirely determined by the ratio of the \acronym{DM} energy and the isotropic exchange energy.

\subsection{Morin transition} 

To determine the Morin transition, we can look at a number of different order parameters. 
In theory, the magnetization $m$ should be zero in the \acronym{afm} phase and assume a finite value in the \acronym{wf} phase. 
However, we can calculate from \eq~\eqref{eq:WeakMagnetisation} that the resulting weak magnetization in the absence of external magnetic fields is only \num{0.000669} (normalized to the saturation value), or $\num{0.00283} \mu_\text{B}$ per iron atom, and therefore too small to be visible in the simulation data without averaging over overly large systems or long simulation times. 

Instead, we focus on the magnitude of the in-plane and out-of-plane components of the Néel vector ($n_z$ and $n_\perp$). 
Since the Morin transition is connected with a reorientation of spins from the $c$ axis into the basal plane, the transition should be clearly visible in these two parameters. 
As shown in \fig~\ref{fig:MorinHysteresis}, we observe the Morin transition at $T_\text{M} = \SI{240(12)}{\kelvin}$, just slightly lower than the experimentally found value of \SI{255}{\kelvin}~\cite{Lebrun2019}.

\subsection{Spin-flop transition} 

\begin{figure} 
\begin{center} 
\includegraphics{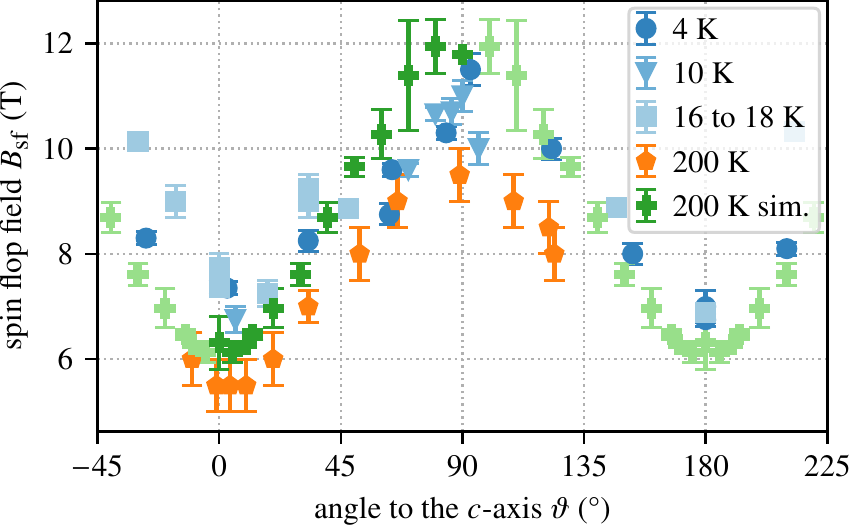} 
\end{center} 
\caption{
Spin-flop field $B_\text{sf}$ in dependence of the angle $\theta$ between the magnetic field and the crystal's symmetry axis. 
Shown are experimental measurements at various low temperatures and at \SI{200}{\kelvin} as well as simulation results at \SI{200}{\kelvin}. 
The experimental data at \SI{200}{\kelvin} were taken from earlier measurements published in Ref.~\cite{Lebrun2019}. 
For the simulation data, the darker points represent the original data and the lighter points are copies of those data points based on symmetry considerations. 
} 
\label{fig:AnisotropyFit} 
\end{figure}

\FFig~\ref{fig:AnisotropyFit} shows measurements of the angle-dependent spin-flop field $B_\text{sf}(\theta)$ both at low temperatures close to zero and at \SI{200}{\kelvin}. 
We expect that the spin-flop fields are largest at low temperatures and then decrease toward $T_\text{M}$. 
And indeed, the measured low-temperature spin-flop fields are clearly higher than at \SI{200}{\kelvin}, but only by a small amount. 
This indicates that the spin-flop fields remain largely constant over this temperature range and only drop off close to $T_\text{M}$, a behavior that has been observed before~\cite{Lebrun2019}. 

Simulation results for $B_\text{sf}(\theta)$ at \SI{200}{\kelvin} are also shown in \fig~\ref{fig:AnisotropyFit} for comparison. 
These results are generally in good agreement, although they slightly overestimate the transition field. 

Simulation results for temperatures close to zero are not shown in \fig~\ref{fig:AnisotropyFit}, because here the critical fields range approximately from \SI{15}{\tesla} to \SI{95}{\tesla}. 
So while the theoretical model agrees well with experimentally measured spin-flop fields at higher temperatures, it overestimates the spin-flop field at low temperatures roughly by a factor of two in the longitudinal case ($\theta = \SI{0}{\degree}$), and even more for transversal fields.

\section{Quantum effects} \label{sec:QuantumEffects} 

\begin{figure} 
\begin{center} 
\includegraphics{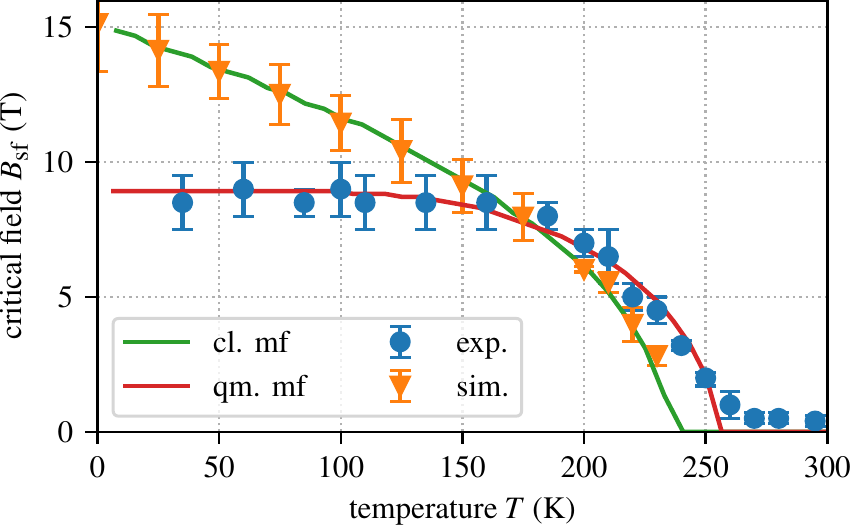} 
\end{center} 
\caption{Comparison of the temperature-dependent spin-flop fields measured and simulated at $\theta = \SI{0}{\degree}$ with the rescaled theoretical predictions from a classical (cl.) and quantum (qm.) mean-field model. 
The experimental data were taken from earlier measurements published in Ref.~\cite{Lebrun2019}. 
} 
\label{fig:QuantumMF} 
\end{figure}

To understand the reason for the large discrepancy of the critical fields at low temperatures, we look at measurements of the longitudinal spin-flop field as a function of temperature, which is shown in \fig~\ref{fig:QuantumMF}. 
As the temperature in the experiment is decreased below the Morin temperature, the critical field rises at first but then reaches a plateau below approximately \SI{150}{\kelvin}. 
This behavior cannot be reproduced by the classical spin model, in which the critical field first rises in line with experiments at temperatures close to $T_\text{M}$ but then continues to increase linearly and hence overestimates the spin-flop field at $T = 0$. 

To ascertain the influence of quantum effects on the temperature dependence of the spin-flop field, we compute the spin-flop field within a mean-field approximation using both a classical model and a quantum model with spin quantum number $S = 2$ (for details, see App.~\ref{app:Meanfield}). 
The mean-field models do not provide quantitative accuracy but they offer a good qualitative picture of the expected shape of the curve. 
For a direct comparison with our data, we therefore rescale the resulting mean-field curves to match the respective Morin temperatures and spin-flop fields, see \fig~\ref{fig:QuantumMF}. 
That way it becomes apparent that 
only the quantum-model curve can be brought into agreement with experimental data for lower temperatures. 
On the other hand, the classical mean-field curve follows the simulation's behavior. 
Above \SI{150}{\kelvin}, the classical model is in reasonable agreement with the experiment.

\section{Conclusions} 

We have presented \fremdwort{ab initio} calculations of the exchange interactions in hematite and how they can be used to parameterize an atomistic spin model that correctly reproduces this complex material's magnetic phases and phase transitions. 
In addition to isotropic exchange and Dzyaloshinskii--Moriya interactions, our simulations incorporate the competing effects of second- and fourth-order on-site anisotropies as well as relativistic and dipolar two-ion anisotropies. 

We have validated our model through comparisons with experimental measurements on a hematite single-crystal. 
Once the anisotropy constants are fitted to the material, we find good quantitative agreement of the Néel and Morin temperatures as well as the weak ferromagnetic canting angle predicted by our model and measured in experiments. 

At low temperatures, deviations between the classical model and experimental results are expected and can be observed. 
Through mean-field approximations, we demonstrate the qualitative differences between a classical and quantum model. 
At low temperatures, only the quantum nature of the thermal fluctuations can explain the temperature dependence of the spin-flop field satisfactorily. 
This allows us to delineate the temperature range in which a classical model is applicable and 
elucidate the deviations arising from quantum effects.

\begin{acknowledgments} 
This work has been supported by the German Research Foundation (DFG) under project No.\ 423441604. 
TD and UN acknowledge additional support from the DFG through project No.\ 425217212 (SFB 1432). 
AD and LS acknowledge support by the National Research, Development and Innovation (NRDI) Office of Hungary under Grant No.\ K131938, PD134579 and TKP2021-NVA-02. 
Computing resources for the \fremdwort{ab initio} calculations were provided by the Governmental Information Technology Development Agency's (KIFÜ) cluster in Debrecen, Hungary. 
The team in Mainz acknowledges additional support from the DFG under project No.~268565370 (SPIN+X, projects A01 and B02) and from the Horizon 2020 Framework Programme of the European Commission under FET-Open grant agreement No.\ 863155 (s-Nebula) and from the Horizon Europe Framework programme under grant agreement No.\ 101070287 (SWAN-on-chip). 
LR gratefully acknowledges support by the Young Scholar Fund at the University of Konstanz and NRDI Office of Hungary under Grant No.\ FK142601. 
\end{acknowledgments}

\appendix 

\section{Simulation of spin-flop fields\label{app:SpinFlopSims}} 

\begin{figure} 
\begin{center} 
\includegraphics{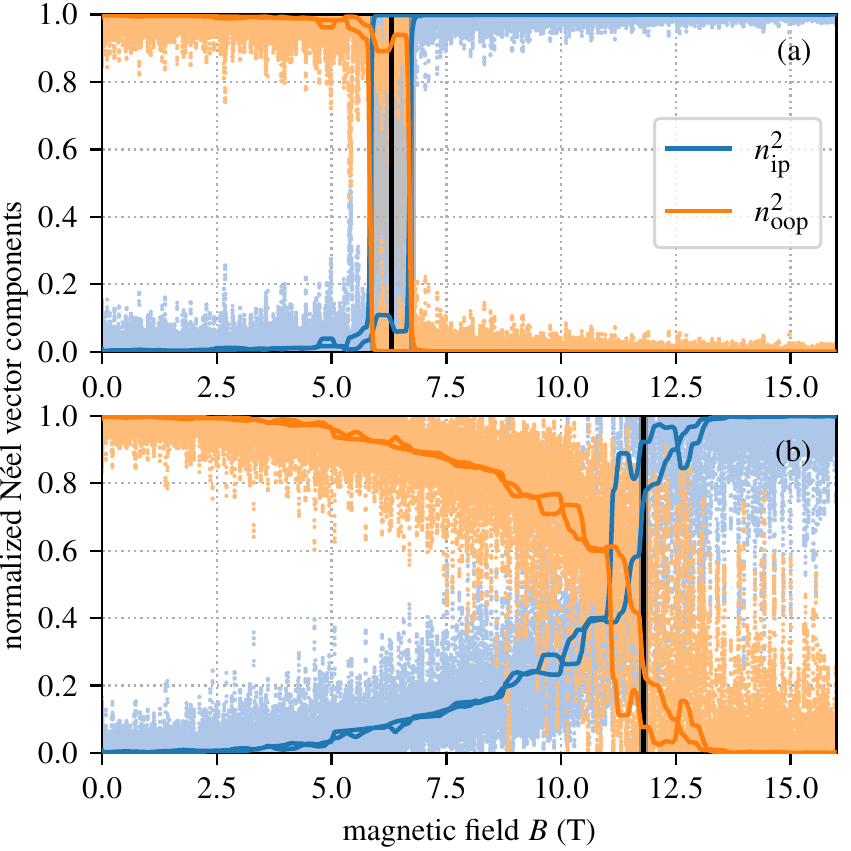} 
\end{center} 
\caption{Simulations of the spin-flop transition at $T = \SI{200}{\kelvin}$ at a magnetic field angle of (a) $\theta = \SI{0}{\degree}$ and (b) $\theta = \SI{90}{\degree}$. 
Plotted are the squared in-plane (ip) and out-of-plane (oop) components of the Néel vector. 
The black lines mark the determined spin-flop field $B_\text{sf}$. 
} 
\label{fig:SpinFlopSimExamples} 
\end{figure}

To determine the spin-flop field $B_\text{sf}(\theta, T)$ in a simulation, the system is first initialized in its ground state at the temperature $T$. 
Then a magnetic field is applied at an angle $\theta$ to the $c$ axis. 
The magnitude of the field is steadily increased up to a certain maximum field strength (chosen to be above the highest expected $B_\text{sf}$) and then decreased again until it reaches zero. 
\FFig~\ref{fig:SpinFlopSimExamples} shows two example simulations. 
It is important to look at both increasing and decreasing fields because a hysteretic behavior can be observed in many cases (see, e.g., \fig~\ref{fig:SpinFlopSimExamples}a). 
We then take the value of $B_\text{sf}$ as the mean between the values determined for increasing and decreasing magnetic field. 
The uncertainty is calculated as the empirical standard deviation of the two values.

\section{Mean-field calculations\label{app:Meanfield}} 

We consider the Hamiltonian 
\begin{equation} 
  H 
  = -\frac{1}{2} \! \sum_{i \neq j,r,s} \!\! \vec{S}_{ir} \tensorial{J}_{ij}^{rs} \vec{S}_{js} 
  - \sum_{i,r} \tensorial{K}_{i}^{r} \left(\vec{S}_{ir}\right)-\sum_{i,r}\mu_{r}\vec{B}\vec{S}_{ir} \, ,\label{eqn1} 
\end{equation} 
where $i,j$ and $r,s$ denote site and sublattice indices, respectively. $\tensorial{J}_{ij}^{rs}$ is the exchange tensor between the spins, $\tensorial{K}_{i}^{r}$ is the on-site anisotropy function containing second-order and fourth-order terms, while $\vec{B}$ is the external field coupling to the spin through the magnetic moment $\mu_{r}$. $\vec{S}_{ir}$ is a unit vector in the classical case and the spin operator with quantum number $S$ in the quantum case, where $\mu_{r}=g\mu_{\textrm{B}}$ is set with $g$ the gyromagnetic factor and $\mu_{\textrm{B}}$ the Bohr magneton. 

In mean-field theory, the expectation values $\langle\vec{S}_{ir}\rangle=\langle\vec{S}_{r}\rangle$ are introduced, which are assumed to depend on the sublattice but not the site. The spin operators are replaced by $\vec{S}_{ir}=\langle\vec{S}_{r}\rangle+\Delta\vec{S}_{ir}$, and the Hamiltonian is approximated such that all terms containing products of the spin fluctuations $\Delta\vec{S}_{ir}$ at different sites are neglected. This results in 
\begin{align} 
  H_{\textrm{MF}} 
  &= \frac{1}{2}\sum_{r,s} \langle\vec{S}_{r}\rangle \sum_{i\neq j} \tensorial{J}_{ij}^{rs} \langle\vec{S}_{s}\rangle 
  \nonumber\\ 
-& \sum_{i,r} \left[\vec{S}_{ir}\frac{1}{2}\sum_{j} \tensorial{J}_{ij}^{rs} \langle\vec{S}_{s}\rangle + \tensorial{K}_{i}^{r} \left(\vec{S}_{ir}\right)+\mu_{r}\vec{B}\vec{S}_{ir}\right] \, .\label{eqn2} 
\end{align} 
Since Eq.~\eqref{eqn2} is a sum of single-particle Hamiltonians, the free energy per unit cell at inverse temperature $\beta=\left(k_{\textrm{B}}T\right)^{-1}$ may be calculated as a sum over the sites, 
\begin{align} 
F_{\textrm{MF}}&=\frac{1}{2}\sum_{r,s} \langle\vec{S}_{r}\rangle \tensorial{J}^{rs} \langle\vec{S}_{s}\rangle\nonumber\\ 
-&\frac{1}{\beta N_{\textrm{c}}}\sum_{i,r}\textrm{ln}\tr e^{-\beta\left(\vec{S}_{ir}\frac{1}{2}\sum_{s} \tensorial{J}^{rs} \langle\vec{S}_{s}\rangle + \tensorial{K}_{i}^{r} \left(\vec{S}_{ir}\right)+\mu_{r}\vec{B}\vec{S}_{ir}\right)}\, ,\label{eqn3} 
\end{align} 
where $N_{\textrm{c}}$ denotes the number of unit cells and $\tr $ denotes an integral over the unit sphere representing the possible spin directions $\vec{S}_{ir}$ in the classical case and the trace in a single-particle basis in the quantum case. 
We introduced the notation $\tensorial{J}^{rs}=\sum_{j} \tensorial{J}_{ij}^{rs}$, which only depends on the sublattice indices due to translational invariance. 

The parameters $\langle\vec{S}_{r}\rangle$ are unknown at this point, and they must be determined such a way that they minimize the mean-field free energy in Eq.~\eqref{eqn3}. Taking the derivative of $F_{\textrm{MF}}$ with respect to $\langle\vec{S}_{r}\rangle$ and setting it to zero leads to the system of mean-field equations 
\begin{equation} 
\langle\vec{S}_{r}\rangle=\frac{\tr \vec{S}_{ir} e^{-\beta\left(\vec{S}_{ir}\frac{1}{2}\sum_{s} \tensorial{J}^{rs} \langle\vec{S}_{s}\rangle + \tensorial{K}_{i}^{r} \left(\vec{S}_{ir}\right) + \mu_{r}\vec{B}\vec{S}_{ir}\right)}}{\tr e^{-\beta\left(\vec{S}_{ir}\frac{1}{2}\sum_{s} \tensorial{J}^{rs} \langle\vec{S}_{s}\rangle + \tensorial{K}_{i}^{r} \left(\vec{S}_{ir}\right) + \mu_{r}\vec{B}\vec{S}_{ir}\right)}}.\label{eqn4} 
\end{equation} 
Note that the right-hand side of Eq.~\eqref{eqn4} indeed defines the expectation value of $\vec{S}_{ir}$ in the single-particle Hamiltonian $H_{\textrm{MF}}$ if the $\langle\vec{S}_{r}\rangle$ values are fixed. However, the meaning of Eq.~\eqref{eqn4} is that the $\langle\vec{S}_{r}\rangle$ parameters have to be determined from it self-consistently to determine the optimal average spin configuration in the mean-field approximation. Since Eq.~\eqref{eqn4} typically has multiple solutions, the real minimum has to be found by substituting these solutions back into Eq.~\eqref{eqn3}; this is a sufficient condition for finding the minimum since the space of the $\langle\vec{S}_{r}\rangle$ parameters is compact. 

Hematite consists of four sublattices, but on the level of the sublattice exchange matrices $\tensorial{J}^{rs}$ the $A$ and $D$ respectively $B$ and $C$ sublattices are equivalent. Since the mean-field equations possess this symmetry, it can be assumed that the solutions satisfy $\langle\vec{S}_{A}\rangle=\langle\vec{S}_{D}\rangle=\langle\vec{S}_{1}\rangle$ and $\langle\vec{S}_{B}\rangle=\langle\vec{S}_{C}\rangle=\langle\vec{S}_{2}\rangle$. Therefore, it is sufficient to treat the two effective sublattices $1$ and $2$ with the interaction tensors 
\begin{align} 
\tensorial{J}^{11}&=\frac{1}{2}\left(\tensorial{J}^{AA}+\tensorial{J}^{AD}+\tensorial{J}^{DA}+\tensorial{J}^{DD}\right) \nonumber\\ 
&=\left[\begin{array}{ccc}J & 0 & 0 \\ 0 & J & 0 \\ 0 & 0 & J+\Delta J\end{array}\right],\label{eqn5} 
\\ 
\tensorial{J}^{12}&=\frac{1}{2}\left(\tensorial{J}^{AB}+\tensorial{J}^{AC}+\tensorial{J}^{DB}+\tensorial{J}^{DC}\right)\nonumber\\ 
&=\left[\begin{array}{ccc}J' & D & 0 \\ -D & J' & 0 \\ 0 & 0 & J'+\Delta J'\end{array}\right],\label{eqn6} 
\\ 
\tensorial{J}^{21}&=\frac{1}{2}\left(\tensorial{J}^{BA}+\tensorial{J}^{BD}+\tensorial{J}^{CA}+\tensorial{J}^{CD}\right)=\left(\tensorial{J}^{12}\right)^{T},\label{eqn7} 
\\ 
\tensorial{J}^{22}&=\frac{1}{2}\left(\tensorial{J}^{BB}+\tensorial{J}^{BC}+\tensorial{J}^{CB}+\tensorial{J}^{CC}\right)=\tensorial{J}^{11}.\label{eqn8} 
\end{align} 
The form of the sublattice interaction tensors described by the parameters $J, \Delta J, J', \Delta J'$ and $D$ is dictated by the system's symmetry. The anisotropy functions are $\tensorial{K}_{i}^{r} \left(\vec{S}_{ir}\right)=d_{2}S_{i,z}^{2}+d_{4}S_{i,z}^{4}$ for $r=1,2$. In the quantum case, we chose the spin quantum number $S=2$, which would result in a magnetic moment of $4\mu_{\textrm{B}}$ giving the closest agreement with the value determined from the \acronym{SKKR} method in \tab~\ref{tab:ab_initio_moments}. This is also the lowest quantum number for which the fourth-order anisotropy can be interpreted; for lower $S$ values, $S_{i,z}^{4}$ may be expressed by $S_{i,z}^{2}$ and constant terms. The expectation values were calculated using a Lebedev--Laikov integration grid~\cite{LebedevLaikov1999Translation} of order $41$ on the unit sphere in the classical case and in the standard basis of the eigenstates of $S_{z}$ for $S=2$ in the quantum case. 

For the magnetic field oriented along the $c$ axis, we considered three different types of solutions of Eq.~\eqref{eqn4}. The first one is $\langle\vec{S}_{1}\rangle=m_{1}\vec{e}_{z}$ and $\langle\vec{S}_{2}\rangle=-m_{2}\vec{e}_{z}$ describes the antiferromagnetic state, with $m_{1}>m_{2}>0$. The second one is $\langle\vec{S}_{1}\rangle=\left(m_{x},m_{y},m_{z}\right)$ and $\langle\vec{S}_{1}\rangle=\left(m_{x},-m_{y},m_{z}\right)$, corresponding to the spin-flop or weak ferromagnetic phase. The third configuration is the paramagnetic one, with $\langle\vec{S}_{1}\rangle=\langle\vec{S}_{2}\rangle=m\vec{e}_{z}$. Phase transitions were detected at the temperature and field values where the minimum of the free energy in Eq.~\eqref{eqn3} switches from the antiferromagnetic first to the spin-flop, then to the paramagnetic configuration.

\end{document}